# Learning to Recommend Third-Party Library Migration Opportunities at the API Level

Hussein Alrubaye*, Mohamed Wiem Mkaouer*, Igor Khokhlov*, Leon Reznik*, Ali Ouni[†], Jason Mcgoff[‡]
*Software Engineering Department, Rochester Institute of Technology, NY, USA
[†]ETS Montréal, University of Quebec, Montréal, QC, Canada
[‡]Xerox Corporation
{hat6622,mwmvse,xk8996,lr}@rit.edu*, ali.ouni@etsmtl.ca[†], Jason.Mcgoff@xerox.com[‡]

*Abstract*—The manual migration between different third-party libraries represents a challenge for software developers. Developers typically need to explore both libraries Application Programming Interfaces, along with reading their documentation, in order to locate the suitable mappings between replacing and replaced methods. In this paper, we introduce RAPIM, a novel machine learning approach that recommends mappings between methods from two different libraries. Our model learns from previous migrations, manually performed in mined software systems, and extracts a set of features related to the similarity between method signatures and method textual documentations. We evaluate our model using 8 popular migrations, collected from 57,447 open-source Java projects. Results show that RAPIM is able to recommend relevant library API mappings with an average accuracy score of 87%. Finally, we provide the community with an API recommendation web service that could be used to support the migration process.

## I. Introduction and Motivation

Modern software systems rely heavily on third-party libraries as a means to save time, reduce implementation costs, and increase software quality while offering rich, robust, and up-to-date features [1], [2], [3]. However, as software systems evolve rapidly, there is a need for appropriate tools, reliable, and efficient techniques to provide developers with support for decision making when replacing their old and obsolete libraries with up-to-date ones. This process of replacing a library with a different one, while preserving the same code behavior, is known as *library migration* [4], [5].

The migration process between libraries is widely acknowledged to be a hard, error-prone, and time-consuming process [6], [2], [3], [1]. Hence, developers have to explore the new library's API and its associated documentation in order to locate the right API method(s) to replace in the current implementation that belongs to the retired library's API. Developers need often to spend significant time to verify that the newly adopted features do not introduce any regression. For instance, previous works have shown that developers typically spend up to 42 days to migrate between libraries [7].

Typically, software development companies tend to assign migration tasks to developers who have more experience to reduce regression risks. For instance, Figure 1 shows that developers who have more than ten years of experience are expected to perform migration more often than a new developer with less than five years of experience. The figure is based on a previous migration benchmark by Alrubaye et al. [6] which contains information about the developers who have performed migration tasks previously, such as, developer names, emails, years of experience, and migration dates. Furthermore, we find that 95.3% of 57,447 Java projects use at least one third party library (APIs). On average, 65 process of API upgrade or migration per project.

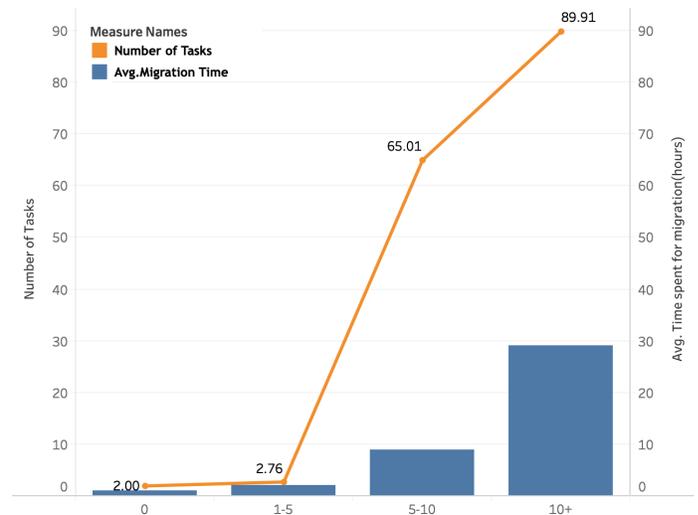

Figure 1: Migration period, clustered by developers years of experience.

A number of migration approaches and techniques have been proposed recently with the aim of identifying what the replacements of a deprecated API are with a newer version of the same API [8], [9], [10], [11]. Other studies recommend which library to adopt, when, retiring another one [12], [13], [14], [15], [16]. However, such approaches do not provide guidance to software developers on how to concretely perform a fine-grained migration at the method-level. Indeed, method-level recommendations have been the focus of many studies, but, only for recommending the same library, across different programming languages or operating systems

[17], [18], [19]. Obviously, there is a need for a more comprehensive recommendation technique that is both library and language independent *i.e.*, it takes as input two different libraries and provides mappings on how to replace one with another at the method level.

In this paper, we introduce a novel machine learning model, labeled as *RAPIM* (Recommending API Migrations), that learns from previously performed migration changes by developers and recommends API-level migrations for similar migration contexts. RAPIM takes as input two different libraries and identifies as output potential mappings between their API methods. The basic idea behind RAPIM is to reuse and take advantage of the valuable migration *knowledge* available in previous manually performed migrations by developers in a different open-source project, *i.e.*, learn from the *"wisdom of the crowd"*. RAPIM uses predefined features related to the similarity of method signatures and their corresponding API documentation to build its model. The model treats the matching game between two API methods as a classification problem, where, for each method from the *retired* API, RAPIM recommends the most relevant method from the *new* API, based on how *close* they are from a lexical and descriptive standpoint. We challenge RAPIM in terms of recommending mappings between various APIs. On average, the model's accuracy was 86.97%. We also challenge the stability of RAPIM with respect to the training size, *i.e.*, we found that the used dataset is sufficient to generalize the model and deploy it.

This study makes the following contributions:

**1.** We propose, RAPIM, an automated approach for library APIs migration that takes as input, two different third-party libraries along with their APIs and documentation and recommends existing mappings between their API methods. RAPIM learns from existing library migration changes manually performed by developers in different open-source projects, then builds a model using various features related to method signatures and method documentation in order to recommend mappings between methods in similar contexts.

**2.** We conduct an empirical study to evaluate RAPIM's performance in detecting mappings for 8 popular migrations, along with comparing it to adapted state-of-the-art migration techniques. Findings show that RAPIM effectively generates correct mappings while improving the state-of-the-art results by 39.51% in terms of accuracy.

**3.** We implement RAPIM and deploy it as a lightweight Web service that is publicly available for software engineers and practitioners to support them in any migration process. We also publicly provide RAPIM's dataset online for replication and extension purposes[1].

## II. Related Work

This section discusses the literature relevant to this work. Several recent studies proposed different API recommendation techniques based on the context of usage. Most of the API recommendation techniques are based on results returned by web search engines and crowd-sourcing, as well as the recommendation of relevant functions, was the focus of multiple studies [20], [21]. McMillan et al. [22] proposed an approach named as Portfolio, a search engine that models the developer's behavior then looks for relevant functions based on (i) call graph similarity and (ii) querying open-source projects using natural language processing. Zhong et al. [12] proposed another approach called MAPO to select API usage patterns and then extracts common sequences that can be used to transform code snippets and make recommendations automatically. CLAN was introduced by McMillan et al. [23] and based on calculating method APIs behavioral similarity by comparing API call-graphs. Software libraries recommendation has been recently formulated as an optimization problem by Ouni et al. [14] using multi-objective search based on NSGA-II [24] to find the best trade-off between maximizing the coverage and similarity between libraries while reducing the number of recommended libraries.

Pandita et al. [18] recommend API mapping between C# and Java using the same API, different programming languages. He detects method mappings between a given source and a target library by automatically discovering possible method mappings across their APIs, using text mining on the functions textual descriptions. Their work was extended to include temporal constraints [19] and to compare text mining between various IR techniques. A dynamic analysis was also used by Gokhale et al. [17] to develop a technique to infer possible mappings between the APIs of Java2 Mobile Edition and Android graphics. The main difference between the existing approaches and our approach, RAPIM, is that they tackle the problem of mapping between methods across different languages, whereas our approach recommends API mappings between different libraries belonging to the same programming language.

## III. Methodology

In this section, we first give an overview of our approach. Then, we detail the different steps and features needed to design our model.

A migration rule is denoted by a pair of a source (removed) library $L_s$ and a target (added) library $L_t$, and represented by $L_s \rightarrow L_t$. For example, *easymock → mockito* represents a *Migration Rule* where the library easymock[2] is migrated to the new library mockito[3]. For a given migration rule $L_s \rightarrow L_t$, let $L_s = m_s^{(i)}$ denote a set of methods that belong to $L_s$, where $m_s^{(i)} = \{m_1, m_2, ..., m_{L_s}\}$, and $L_t = m_t^{(i)}$ denotes a set of methods that belong to $L_t$, where $m_t^{(i)} = \{m_1, m_2, ..., m_{L_t}\}$. Our goal is to find an alignment between both $L_s$ and $L_t$.

---

[1] http://migrationlab.net/index.php?cf=icsme2019

[2] http://easymock.org
[3] https://site.mockito.org

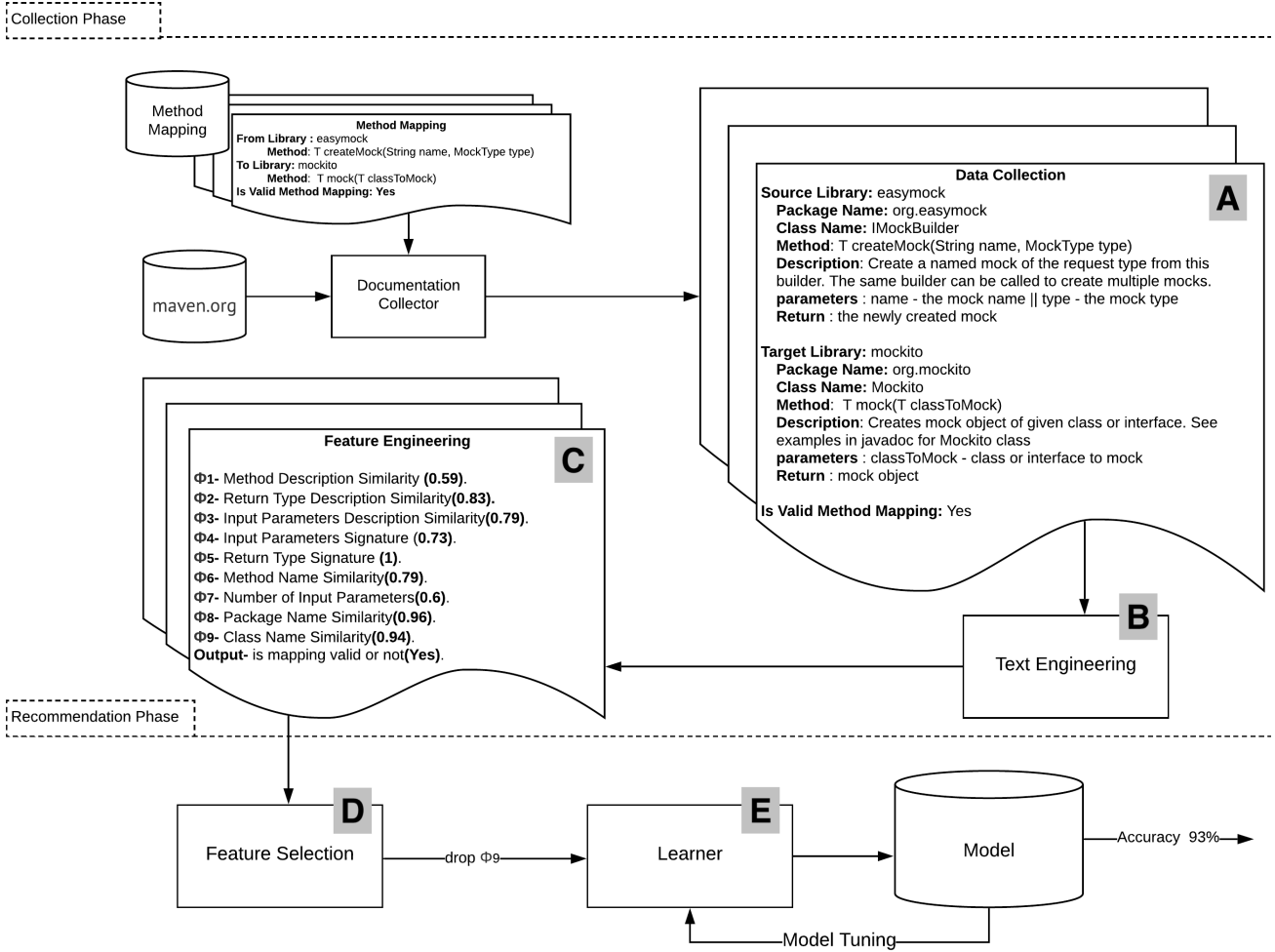

Figure 2: The proposed RAPIM approach for method APIs mapping recommendation.

$$f : L_s \rightarrow L_t \qquad (1)$$

in such a way that each source method $m_s^{(i)} \in L_s$ is mapped to an equivalent target method $m_t^{(i)} \in L_t$, this process is called *Method Mapping*.

Figure 2 provides an overview of RAPIM approach which consists of two main phases: the first phase, called (1) *Collection Phase*, collects the necessary information, e.g., library documentation, for all the mappings contained in the data set [6], to generate the features. This phase starts with (A) the collection of APIs and their corresponding documentation; (B) text preprocessing, and (C) the feature engineering that used to extract the feature from method signature and API documentation. The second phase, called (2) *Recommendation Phase*, starts with (D) the selection of relevant features, before (E) passing them to the learner. The learner generates RAPIM model that used to recommend relevant library API mappings between two libraries. In the following, we detail RAPIM's five main processes.

*A. Data Collection*

This phase takes two inputs. The first input, method mappings, consists of a manually inspected dataset of valid and invalid method mappings for different migration rules from a study by Alrubaye et al. [6]. For example, in Figure 2, for a given migration rule *easymock* → *mockito*, we identify one of the valid method mappings between the two following methods *createMock(String name, MockType type)* → *mock(T classToMock)*.

The second input of this phase is the API documentation, which is represented by the *Documentation*

*Collector*[4] in Figure 2. For a given method mapping, the *Documentation Collector* collects the API documentation for both the source method and the target method. Based on a migration rule, it automatically downloads the library documentation as a *jar* file for all library releases involved in migrations. Our approach relies on the libraries documentation on Maven Central Repository[5]. The Documentation Collector then converts the API documentation from a *jar* file to multiple *HTML* source files using the *doclet API* [6]. It parses all of the *HTML* files and collects the documentation related to class descriptions, method descriptions, parameter descriptions, return descriptions, package names, and class names. The Documentation Collector identifies the documentation associated with every method mapping. The collection process ends when all the information associated with every method involved in all method mapping in the dataset are collected.

*B. Text Engineering*

Our approach aims at automatically recommending API method mappings to support developers in their library migration tasks. The migration task involves typically the analysis of structured and unstructured data sources, including method signatures, textual API descriptions, code snippet examples from open-source repositories, etc. To automatically explore such data, we deploy information retrievals (IR) techniques such as text preprocessing, vector space model, and cosine similarity to preprocess our sources.

*1) Information Extraction (IE):* Let $d$ be method signature (name, class name, or package name). In this step, we extract $d^*$ using the function named Information Extraction *IE* as follows:

$$d^* = IE(d) \qquad (2)$$

For example, in Figure 2, if $d$ is the target API package name, then $d^*$ is generated using *IE* which is described as follows:

---
**Information Extraction (*IE*)**

**input (*d*)**: *'com.IMockBuilder'*.
**1- Special Characters Cleanup:** In this step, we replace all special characters such as dots with a space. For the given input, the output for this step is $'com <space> IMockBuilder'$.
**2- Camel Case Splitter:** In this step, we split all identifiers with using camel case. The output for this step is $'com <space> I <space> Mock <space> Builder'$.
**Output(*d\**):** *'com I Mock Builder'*

---

*2) Text Preprocessing (TPP):* $d$ may have a mix of words and special characters, such as a dot, a colon, etc. In text processing *TPP*, we clean the documents of special characters and common English words, such as "the" and "is". We then apply a stemming transformation to all extracted words to put them in their root format using Natural Language Processing[7] (NLP). This process helps to reduce the noise when calculating the similarity between two documents.

$$\widehat{d} = TPP(d) \qquad (3)$$

For example, in Figure 2, if $d$ is a source method description, then $\widehat{d}$ is generated, using *TPP*. The *TPP* process is described as follows:

---
**Text Preprocessing (*TPP*)**

**input (*d*)**: 'Create a named mock of the request type from this builder. The same builder can be called to create multiple mocks.'
**1- Tokenization:** In this step, we convert text words into a list of tokens so that we can process each token alone.
['Create', 'a', 'named', 'mock', 'of', 'the', 'request', 'type', 'from', 'this', 'builder', '.', 'The', 'same', 'builder', 'can', 'be', 'called', 'to', 'create', 'multiple', 'mocks', '.']
**2- Unnecessary punctuation removal:** This is the process of removing unnecessary punctuation, tags such as '.' from a list of tokens.
['Create', 'a', 'named', 'mock', 'of', 'the', 'request', 'type', 'from', 'this', 'builder', 'The', 'same', 'builder', 'can', 'be', 'called', 'to', 'create', 'multiple', 'mocks']
**3- Stop and reserved words removal:** In this step, we remove all English words and reserved words[a] such as "a", "of", "the", "from", "this", "can", "be", "to".
['Create', 'named', 'mock', 'request', 'type', 'builder', 'builder', 'called', 'create', 'multiple', 'mocks']
**4- Lemmatization:** is the process of reducing words to the root, This helps to remove inflection and reduce inflectional forms. For example *called, calling, call's,* ⇒ *call, mocks,* ⇒ *mock*.
['Create', 'name', 'mock', 'request', 'type', 'builder', 'builder', 'call', 'create', 'multiple', 'mock']
**5- Ouput($\widehat{d}$):** This last step is to convert all characters to lowercase and combine all tokens into one string.
*'create mock request type builder builder create multiple mock'*

[a]http://www.textfixer.com/resources/common-english-words.tx

---

[4]http://migrationlab.net/tools.php?cf=icsme2019&tool=DoC
[5]central.maven.org
[6]https://goo.gl/S3xRwk

[7]nlp.stanford.edu/software/corenlp.shtm

*3) Vector Space Representation:* As part of generating the features, we calculate the similarity between the source method documentation *s* and the target method documentation *t*. This includes the similarity between each method description, method name, or method return type description of *s*, and *t*. To calculate the similarity between two textual documents, we first need to convert the text to a numeric vector and then calculate their closeness using cosine similarity. To convert the text into a numeric vector, we use the Term Frequency-Inverse Document Frequency (TF-IDF) technique. For a given document, the weight vector $W_d$ represents an array of frequency weights for each term in the document. The weight for each term $w_{t,d}$ is based on the classic $tf * idf$ weighting, as shown in equation 4 where $tf_{t,d}$ is the number of times a term *t* appears in a document and $t_n$ is the number of terms in the document. While *N* is the number of documents. In our case, *N* = 2 since we are performing binary comparisons(source and target method). $df_t$ is the number of documents in which the term *t* has appeared. In our case, it has the value of 1 if it appears in one document or 2 if it appears in both documents.

$$W_d = \begin{bmatrix} w_{t_1,d} \\ w_{t_2,d} \\ \vdots \\ w_{t_n,d} \end{bmatrix}, w_{t,d} = \frac{tf_{t,d}}{t_n} * log\left(\frac{N}{df_t}\right) \quad (4)$$

Then, we use cosine similarity $sim(s,t)$ to measure how similar two vectors are, based on the dot product of their magnitude [25]. For a given source weight vector $W_s$, and a target weight vector $W_t$, we calculate $sim(s,t)$ between the two vectors using the equation 5 which outputs a value between [0-1], where 0 means the two documents are completely distinct, and 1 means both documents are identical. The higher the $sim(s,t)$ is, the closer the two documents are.

$$sim(s,t) = cos(s,t) = \frac{W_s \cdot W_t}{||W_s|| \cdot ||W_t||} \quad (5)$$

*C. Feature Engineering*

The machine learning model needs numeric features to process. All the data that we have so far is text data. In this process, we extract numeric features from the source and target method information that we think may help the machine learning model to recommend more accurate results. Initially we extract nine different features $\varphi_1(s,t)$ to $\varphi_9(s,t)$ from *s* and *t* method information, and one binary class *Output* which is either valid or invalid and predefined in the dataset. Every feature is calculated between every method from the source library $L_s$, with every method from the target library $L_t$.

*1) Method Description $\varphi_1$:* we extract $\varphi_1(s,t)$, by calculating the cosine similarity between the source method description $md_s$, and the target method description $md_t$. We have decided not to apply text preprocessing *TPP* on the methods' description because it could have code examples that will be cleaned if we apply *TPP* on text. We have found that keeping these code examples increases the accuracy by 3% as opposed to removing them using the *TPP* process.

$$\varphi_1(s,t) = sim(md_s, md_t) \quad (6)$$

For instance, to calculate $\varphi_1(s,t)$ from the example in Figure 2, we calculate the cosine similarity between $md_s$ (*"Create a named mock of the request type from this builder. The same builder can be called to create multiple mocks."*), and $md_t$ (*"Creates mock object of given class or interface. See examples in Javadoc for Mockito class"*). In this case, the similarity score is (0.59).

*2) Return Type Description $\varphi_2$:* This feature is extracted by applying *TPP* on the source method return type description $rtd_s$, and the target method return type description $rtd_t$, to generate $\widehat{rtd_s}$, and $\widehat{rtd_t}$. The cosine similarity is then applied between $\widehat{rtd_s}$, and $\widehat{rtd_t}$.

$$\varphi_2(s,t) = sim(\widehat{rtd_s}, \widehat{rtd_t}) \quad (7)$$

For instance, to calculate $\varphi_2(s,t)$ from the example in Figure 2, we apply *TPP* on both $rtd_s$ (*"the newly created mock"*) and $rtd_t$ (*"mock object "*) to get $\widehat{rtd_s}$ and $\widehat{rtd_t}$. We then calculate the cosine similarity between $\widehat{rtd_s}$ and $\widehat{rtd_t}$. In this case the similarity score is (0.83).

*3) Input Parameters Description $\varphi_3$:* This feature is extracted by applying *TPP* on the source method input parameters description $ipd_s$ and the target method input parameters description $ipd_t$ to generate $\widehat{ipd_s}$ and $\widehat{ipd_t}$. We then apply the cosine similarity between $\widehat{ipd_s}$ and $\widehat{ipd_t}$.

$$\varphi_3(s,t) = sim(\widehat{ipd_s}, \widehat{ipd_t}) \quad (8)$$

For instance, to calculate $\varphi_3(s,t)$ from the example in Figure 2, we apply *TPP* on both $ipd_s$ (*"name -the mock name — type - the mock type"*), and $ipd_t$ (*"classToMock - class or interface to mock"*) to get $\widehat{ipd_s}$, and $\widehat{ipd_t}$, then We calculate the cosine similarity between $\widehat{ipd_s}$, and $\widehat{ipd_t}$. In this case the similarity score is (0.79).

*4) Input Parameters Signature $\varphi_4$:* This feature is extracted by applying *IE* on source method input parameters signature $ips_s$, and target method input parameters signature $ips_t$ that generate $ips_s^*$, and $ips_t^*$. Then apply the cosine similarity between $ips_s^*$, and $ips_t^*$.

$$\varphi_4(s,t) = sim(ips_s^*, ips_t^*) \quad (9)$$

For instance, to calculate $\varphi_4(s,t)$ from the example in Figure 2, we apply *IE* on both $ips_s$ (*"String name, MockType type"*), and $ips_t$(*"T classToMock"*) to get $ips_s^*$, and $ips_t^*$, then We calculate the cosine similarity between $ips_s^*$, and $ips_t^*$. In this case the similarity score is (0.73).

*5) Return Type Signature $\varphi_5$:* This feature is extracted by comparing source method return type signature $rts_s$, and target method return type signature $rtd_s$, if they have same return type, we return one otherwise we return zero.

$$\varphi_5(s,t) = \begin{cases} 1 & \text{if } rts_s \text{ is equal to } rts_t \\ 0 & \text{if } rts_s \text{ is not equal to } rts_t \end{cases} \quad (10)$$

For instance, to calculate $\varphi_5(s,t)$ for example in Figure 2, both $rts_s$, and $rts_t$ return generic which is $T$, in this case the result for this matrix will be one (1).

*6) Method Name $\varphi_6$:* This feature is extracted by applying *IE* on source method name $methodName_s$, and target method name $methodName_t$ that generate $methodName_s^*$, and $methodName_t^*$. Then apply the cosine similarity between $methodName_s^*$, and $methodName_t^*$.

$$\varphi_6(s,t) = sim(methodName_s^*, methodName_t^*) \quad (11)$$

For instance, to calculate $\varphi_6(s,t)$ from the example in Figure 2, we apply *IE* on both $methodName_s$ ("createMock"), and $methodName_t$ ("mock") to get $methodName_s^*$, and $methodName_t^*$, then We calculate the cosine similarity between $methodName_s^*$, and $methodName_t^*$. In this case the similarity score is (0.79).

*7) Number of Input Parameters $\varphi_7$:* This feature is extracted by calculating the ratio between number of input parameters in source method $inputParamCount_s$ and number of input parameters in target method $inputParamCount_t$ as shown in equation 12.

$$\varphi_7(s,t) = 1 - \frac{|inputParamCount_s - inputParamCount_t|}{inputParamCount_s + inputParamCount_t} \quad (12)$$

For instance, to calculate $\varphi_7(s,t)$ from the example in Figure 2, we find different between $inputParamCount_s$ which has two parameters which are *name, and type*, and $inputParamCount_t$ that has one input parameters which is *(classToMock)*, so the different is (0.6).

*8) Package Name $\varphi_8$:* This feature is extracted by applying *IE* on source method package name $packageName_s$, and target method package name $packageName_t$ that generate $packageName_s^*$, and $packageName_t^*$. Then apply the cosine similarity between $packageName_s^*$, and $packageName_t^*$.

$$\varphi_8(s,t) = sim(packageName_s^*, packageName_t^*) \quad (13)$$

For instance, to calculate $\varphi_8(s,t)$ from the example in Figure 2, we apply *IE* on both $packageName_s$ ("org.easymock"), and $packageName_t$ ("org.mockito") to get $packageName_s^*$, and $packageName_t^*$, then We calculate the cosine similarity between $packageName_s^*$, and $packageName_t^*$. In this case the similarity score is (0.96).

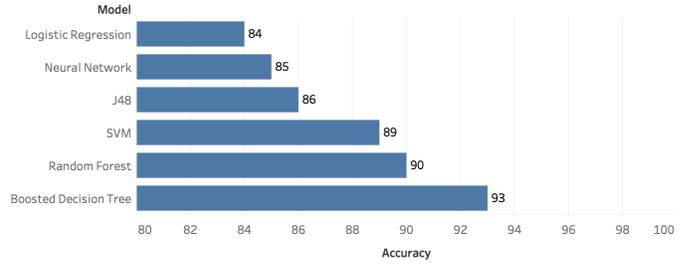

Figure 3: Comparative study between learners, in terms of accuracy.

*9) Class Name $\varphi_9$:* This feature is extracted by applying *IE* on class name where source method lives $className_s$, and class name where target method lives $className_t$ that generate $className_s^*$, and $className_t^*$. Then apply the cosine similarity between $className_s^*$, and $className_t^*$.

$$\varphi_9(s,t) = sim(className_s^*, className_t^*) \quad (14)$$

For instance, to calculate $\varphi_9(s,t)$ from the example in Figure 2, we apply *IE* on both $className_s$ ("IMockBuilder"), and $className_t$ ("Mockito") to get $className_s^*$, and $className_t^*$, then We calculate the cosine similarity between $className_s^*$, and $className_t^*$. In this case the similarity score is (0.94).

*D. Feature selection*

We predefined nine features $\varphi_1$ to $\varphi_9$, however, we are not sure if all of these features are helpful to the learner in recommending better results. We applied *Filter Based Feature Selection* [26] which shows how much each feature contributes to recommending the output. The filter shows us that $\varphi_9$ does not have any contribution in recommending the *output* class. In this case, there are two methods from two different libraries written by two different developers that could have the same class name. So, we drop this feature.

*E. Learner*

There are a number of machine learning algorithms designed precisely for this situation. Such an algorithm takes the form of *classifier* which operates on *instances* [27]. For our purposes, an instance is a feature vector extracted between a source and a target method ( $\varphi_1$ to $\varphi_8$). In the training phase, we feed the classifier a set of instances along with labeled "output". The label output is binary judgment by the previous study [6] that classify the method mapping as "valid" or "invalid". We normalize all the instances using *z-score*, to avoid overfitting problem.

When the training is complete, The classifier generates a *model*. We give a model an instance that has not seen before. The model predicates the probability that it belongs in the valid or invalid method mapping class. We used Azure Machine learning studio [8]

We conduct a comparative study between various potential classifiers, which can be candidates for being used as our base model, as follows:

*Ensemble Learning.* it randomly selects samples from the dataset, and for every sample, it applies a Decision Tree (DT) to build and test the learner using the remaining rest of the dataset. Then, it uses the miss-predicted samples as part of the training dataset used by the next learner. Afterwards, it finds the probability of the output for all learners. This improves the learner's accuracy and reduces the over-fitting problem. After tuning, we found that 233 data-set samples are giving the minimum error as shown in Figure 4 This means that we have 233 possible accepted solutions.

As shown in Figure 3, we compared between various state-of-the-art learners, including, neural networks, Support Vector Machines (SVM), Random Forest and Boosted Decision Trees (BDT). As they require relatively high number of records in the training and the testing set, the logistic regression and neural network models demonstrated the worst results among tested classifiers. The tested neural network models were easily over-fitting which led to an 85% accuracy rate. We also tried various numbers of neurons in the hidden layer, but the results did not improve. A J48 decision tree had similar results to the neural network model, with an 86% accuracy rate. The SVM and random forest classifiers gave respectively an accuracy rate of 89% and 90%. A Two-Class Boosted Decision Tree was the best learner for our dataset with an accuracy rate of 93%. Boosted Decision Trees are known for their good performance on relatively small datasets due to it use of an ensemble of Decision Trees and weighted voting.

## IV. Experimental Design

We design our methodology to answer the following two research questions.

- **RQ1. (Accuracy)** To what extent is RAPIM able to generate the correct method mappings? How does it perform in comparison with the state-of-the-art techniques?
- **RQ2. (Training Size)** What is the minimum training data that RAPIM needs to recommend an optimal mapping?

To answer RQ1, we evaluate the accuracy of RAPIM in recommending correct method mappings for eight popular migrations. To ensure a fair comparison, we perform our comparative study using the same dataset [6] (i.e., input migration rules that run under the same execution environment). RAPIM and Learning-To-Rank use one binary output class and the same exact set of eight features that we have discussed previously. Because these approaches are supervised learning, we split our data-set into training and testing. For every run we have one migration rule per testing set and the remaining go to a training set. RAPIM uses the training set to learn recommendation patterns, while Learning-to-rank uses the training set to compute the weights for the features. TMAP and MS will only consider the input migration rule because they are not learning algorithms.

**Accuracy.** is the ratio of all correctly recommended method mappings divided by all of the correct and incorrect recommended mappings.

$$Accuracy = \frac{Tp + Tn}{Tp + Tn + Fp + Fn}$$

where $Tp$ is the total number of valid mappings that were recommended as a valid mapping. $Fp$ is the total number of invalid mappings that were recommended as a valid mapping. $Tn$ is the total number of invalid mappings that were recommended as an invalid mapping. $Fn$ is the total number of valid mappings that were recommended as an invalid mapping. The higher the *Accuracy* value, the better the recommendation.

**Error.** We use the following equation to measure the tuning error, where a lesser *Error* value will mean the results are better.

$$Error = 1 - Accuracy$$

To answer RQ2, we combine all the mappings from all the rules and then randomly split them into 10 equal folds to mitigate the danger of over-fitting. This allows for the creation of a more diverse set of mappings in each fold. We then run the algorithm nine times. For every run, we increase the training size, decrease the testing size, and measure the *Accuracy*. We start with one fold for training and nine folds for testing. We then increase the folding size for training by one and decrease the folding size for testing by one, and so on, until we have nine folds for training and one fold for testing. The goal of answering this research question is to evaluate the impact of the training data sizes on RAPIM's accuracy. In order to provide the solution as a web-service, we need to make sure that our model has been trained on sufficient data, Therefore, we perform this experiment to verify whether our approach is a stable one when using the existing set of migrations as training.

### A. State of the art approaches

In this section, we describe the implementation of three state-of-the-arts approaches that we have compared with our approach. We adopted these three state-of-the-art approaches to recommend method mapping between $L_s$, and $L_t$. For every method in $L_s$, each approach calculates the similarity score with every method in $L_t$ and returns the method that has the highest matching score at $k = 1$. We selected k=1 because we only recommend one target method for every source method, for all approaches.

[8]https://studio.azureml.net/

*1) Learning to Rank (LTR):* We adopt library recommendation as ranking problem. We use the same features that we extracted in Section 2, along with the dataset as training to calibrate the weights of the features. A score is given for each pair of methods, belonging to the source and target API. The scoring function is a linear combination of features, whose weights are automatically trained on based on the previous mappings. The ranking function is defined as follows:

$$LTR_{score(s,t)} = \sum_{i=1}^{8} W_i^{LTR} * \varphi_i(s,t) \qquad (15)$$

Where each feature $\varphi_i$ measures the specific relationship between the source method $s$ and the target method $t$ of first eight features that discussed in the previous section. The weight parameters $W_i^{LTR}$ are the results of training on the previously solved method mappings. So, for each source method, learning-to-rank ranks the candidate target methods that are most likely to replace it. To ensure the fairness between learning-to-rank and other algorithms under comparison, we only consider the highest ranked method (TOP1).

*2) TMAP:* The Pandita1 [19] approach ranks each method mapping based on the similarity of five features.

$$TMAP_{score(s,t)} = \sum \varphi_1(\hat{s},\hat{t}) + \varphi_6(s,t) + \varphi_8(s,t) + \qquad (16)$$
$$\varphi_9(s,t) + \varphi_x(s,t)$$

Where $\varphi_x(s,t)$ is calculated by applying *TPP* on the source method class description $cd_s$ and the target method class description $cd_t$ that generates $\widehat{cd_s}$ and $\widehat{cd_t}$. We then apply the cosine similarity between $\widehat{cd_s}$ and $\widehat{cd_t}$. While $\varphi_1(\hat{s},\hat{t})$ is calculated by applying *TPP* on the source method description $md_s$ and the target method description $md_t$ that generates $\widehat{md_s}$ and $\widehat{md_t}$. We then apply the cosine similarity between $\widehat{md_s}$ and $\widehat{md_t}$. Other features are generated in the same manner as the previous section.

*3) Method Signature (MS):* This approach calculates the method signature similarity for each combination of methods as follows [28]:

$$MS_{score(s,t)} = 0.25 * sm(rts_s, rts_s) + 0.25 * lcs(ips_s, ips_t) + $$
$$0.5 * lcs(methodName_s, methodName_t) \qquad (17)$$

where $sm()$ calculates the token-level similarity [8] between the two return types and $lcs()$ computes the longest common sub-sequence between the two given input method names [29].

### B. Parameter Tuning

Parameter tuning significantly impacts the performance of the learner for a particular problem [30]. For this reason, we tune the learner in order to improve the accuracy. Since our learner is a *Two-Class Boosted Decision Tree(BDT)*, we start our tuning using the following default inputs: *Maximum Number of leaves=20, Minimum leaf instances=10, Learning rate=0.2*, and *Number of trees=100*. We then iteratively tune the learner until we get a minimum error that cannot be improved upon. Figure 4 shows how the error decreased from 15% to 0.5% after we tuned the Decision Tree inputs. We can see that having the number of trees to 233 has stabilized the error rate at 0.5%. We have concluded that the best values for the learner input parameters are: *Number of leaves=6, Minimum leaf instances=47, Learning rate=0.14*, and *Number of trees=233*.

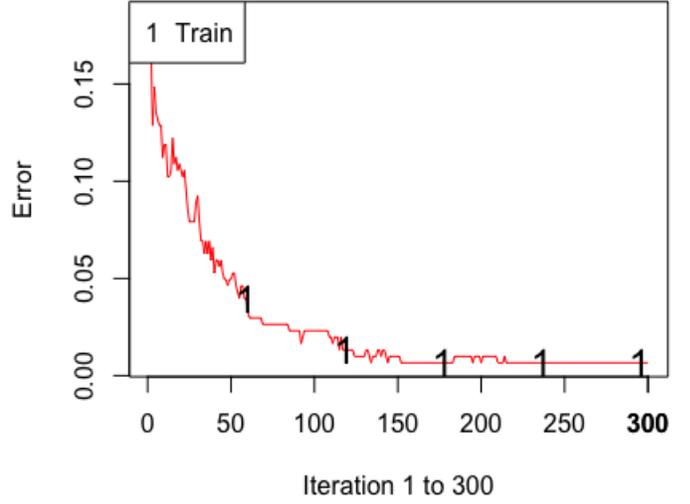

Figure 4: Error with tuning.

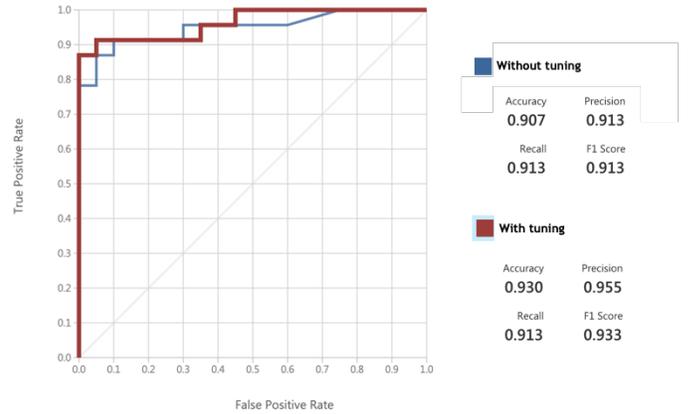

Figure 5: ROC Curve for BDT with and without tuning.

Figure 5 illustrates the comparison of learner recommendations with and without tuning. We see that, with turning the learner is farther from the curve and the accuracy is improved by 3%.

The features weights of LTR also needs to be cal-culated, The LTR parameters' weight is trained on all of the training set except the given migration rule data. The average parameters' weights $W_i^{LTR}$ are the following: $W_1^{LTR}$ = 0.41, $W_2^{LTR}$ = 0.10, $W_3^{LTR}$ = 0.17, $W_4^{LTR}$ = 0.39, $W_5^{LTR}$ = 0.49, $W_6^{LTR}$ = −0.11, $W_7^{LTR}$ = 0.37, $and$ $W_8^{LTR}$ = −0.00058.

Table I: Performance of approaches under test, in terms of accuracy, across 8 migrations.

| Migration Rule | LTR | TMAP | MS | RAPIM |
|---|---|---|---|---|
| logging→slf4j | 28.26% | 21.73% | 26.08% | 85% |
| comm-lang→slf4j | 33.33% | 33.33% | 33.33% | 89.9% |
| easymock→mockito | 26.66% | 46.66% | 46.66% | 80% |
| testng→junit | 51.72% | 51.72% | 37.93% | 98% |
| slf4j→log4j | 77.77% | 66.66% | 77.77% | 85% |
| json→gson | 35.29% | 47.05% | 41.17% | 85% |
| json-simple →gson | 60.0% | 60.0% | 40.0% | 92.9% |
| gson→jackson | 66.66% | 50% | 50% | 80% |
| Average Accuracy | 47.46% | 47.14% | 44.11% | 86.97% |

## V. Results

*A. Results for RQ1.*

We calculated the accuracy of the mappings that are generated by RAPIM, in addition to other state-of-the-art approaches: LTR, TMAP [19], and MS [28].

Table I illustrates the accuracy of the four approaches for eight migration rules. RAPIM has the highest accuracy across all of the rules and it only varies from 80% to 98% on average. We observed that the accuracy score achieved by RAPIM is significantly higher than the three other approaches by 39.51%.

To illustrate how different approaches result in a different levels of accuracy, we qualitatively analyze the results, and we have extracted the following example[9] in Figure 6, which was performed during the migration between *json* and *gson*.

In Figure 6 (A), for a given source method *"String toJSONString()"*, all four approaches were able to recommend the correct target method *"String toString()"*. MS recommends the correct target method because the return type and the input parameters for both methods are the same. Also, the method names are very similar. TMAP recommends the correct target method because both methods have a similar description $\varphi_1$, and name $\varphi_6$. LTR recommends the correct method because both methods have a similar, description $\varphi_1$, input parameter signature $\varphi_4$, and return type $\varphi_5$. These three features also have high weights when compared to other features, which increases the accuracy of the ranking algorithm.

In Figure 6 (B), for a given source method *"JSONObject put(String key, int value)"* only RAPIM was be able to recommend the correct target method *"void addProperty(String property, Number value)"*. LTR recommends *"void addProperty(String property, String value)"* as the target method instead of *"void addProperty(String property, Number value)"*. The reason that LTR recommends the wrong method is because the input parameter for the recommended method *"String value"* has a higher similarity to the source method for $\varphi_3$, and $\varphi_4$ than the similarity of *"Number value"* to the correct target method, while other features have the same values for both target methods. So, this is due to the polymorphic nature of the method. So LTR did recommend the right method name, but not the one with the right types of input parameters. TMAP recommends *"JsonElement parse(JsonReader json)"* as the target method because $\varphi_1$, and $\varphi_9$ have a higher similarity to the recommended target method and source method than the correct target method. MS recommends *"JsonElement get(String memberName)"* because it has a higher signature similarity score to the source method *"put"* than the correct target method has to the source method. In both cases RAPIM recommends the correct mapping because it has learned to detect these types of patterns through its various generated decisions tress.

[9]http://migrationlab.net/redirect.php?cf=icsme2019&p=1

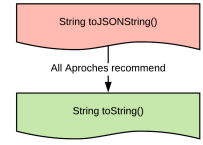
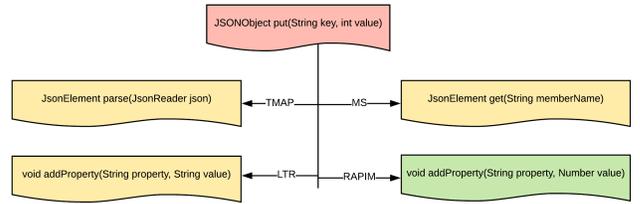

Figure 6: Samples of method mappings between *json* and *gson*.

Through our manual analysis of the results, we notice that, all approaches are generally challenged in the following contexts: *Method Overloading.* It refers to two methods with same name, but with different number of parameters, type of parameters, or the order of the parameters. *Polymorphic Methods.* They are overridden in a class hierarchy where the subclass method has the same name and number of parameters of the base class method, but with different types. *Generic methods* Meth-ods including type parameters for both the returned data and the data that is passed to the method. This allows for the method to operate on objects of various types. There are also harder cases when source and target methods differ in name, return types and even input parameters. Finally, there are also few methods without proper documentation, which also can be a challenge mainly for TMAP, LTR and RAPIM.

> **Summary for RQ1.** The qualitative analysis of 8 migration rules has demonstrated that that RAPIM's accuracy has an average of 86.97%, while, the maximum accuracy scored by the other approaches is 47.46%. Thus, RAPIM has increased the accuracy of the state-of-art approaches by 39.51%.

*B. Results for RQ2.*

Figure 7 shows the performance of RAPIM, in terms of Accuracy, as a function of the number of folds used for training. We observe that by increasing the training size, the accuracy has slightly increased from 83.3% (when trained using one fold) all the way to 92% (when trained with nine folds). We statistically tested the significance of the difference in values by applying the Mann-Whitney U Test and we found no significant difference between the result of training on fold2 and the result of training on all remaining folds.

We confirm that 10% from the training set is sufficient to recommend a method mapping with an accuracy of 83.3% as shown in Figure 7. Also, 30% from the data-set, used as training, was enough to recommend a method mapping with an average accuracy of 86.90%. This argues that the extracted features are independent and that using only a subset of the training set is enough for RAPIM to achieve acceptable accuracy.

> **Summary for RQ2.** RAPIM achieves near optimal accuracy using only a subset of the training set. Thus, using the whole training set raises our confidence that our model, being exported as web-service for practitioners to use, will achieve satisfactory results.

## VI. Threats to validity

We report, in this section, any potential factors that threaten the validity of our analysis.

*A. Construct validity*

Threats to construct validity describe concerns about the relationship between theory and observation and, generally, this type of threat is mainly constituted by any errors related to measurements. For calculating the features, running the experiments, we have used popular frameworks and libraries such as Microsoft AI [31], and NLTK [32]. For the comparative study, we have implemented LTR and TAMP, and this is another threat to validity. We mitigated this threat by verifying that our findings match the results of the previous papers. For instance, LTR's accuracy@K=1 varies between 10% to 45%, while in our study, LTR's accuracy@K=1 is 47%.

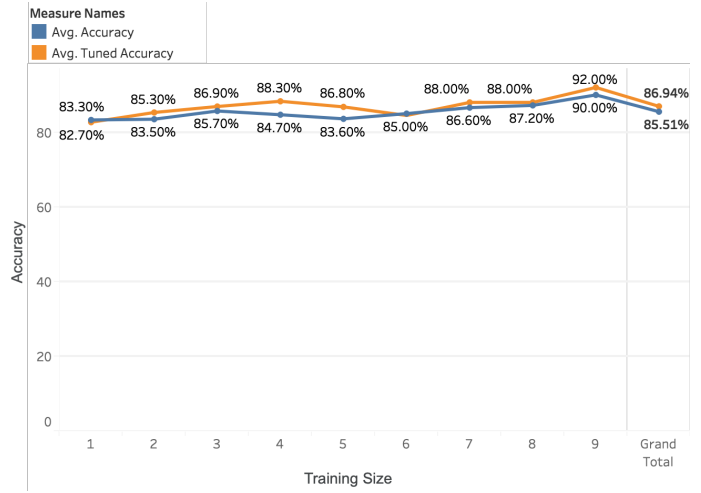

Figure 7: Impact of training data size.

*B. External validity*

Threats to external validity are connected to the generalization of the obtained results. All our tested libraries were Java libraries, belonging to Maven, and so they follow the Object-Oriented principles and Maven naming and documentation conventions, and this may represent a threat to our classification since it heavily depends on textual similarities. Also we should report that not all methods were documented, and this may also impact the performance of some of our features, but these instances were very limited. Since our findings show that our approach did achieve good results across various libraries, written by different developers, even with a few sample of training.

## VII. Conclusion

This study addressed the challenge of recommending method mapping when migrating between third-party libraries. We have described a novel approach that recommends method mappings between two unknown libraries using features extracted from the lexical similarity between method names and from the textual similarity from method documentations. We evaluated our approach by testing how our approach, and three states of art approach's Pandita1 [19], Nguyen [28], and LTR Xin [33] recommend method mappings for 8 given popular library migrations. We find that our approach out performs all existing state of the art approaches. The qualitative and comparative analysis of our experiments indicates that our approach significantly increases the accuracy of recommended mappings by an average accuracy of 39.51%, in comparison with existing state-of-the-art studies. As part of our future investigations, we plan extending the number of migrations used, along with comparing against a larger set of binary classifiers. We also plan on increasing the feature space by including the usage context for methods, in the code.